\documentclass[%
reprint,
superscriptaddress,
showpacs,preprintnumbers,
 amsmath,amssymb,
 aps,
prx,
]{revtex4-1}

\usepackage{graphicx}   
\usepackage{dcolumn}
\usepackage{bm}
\usepackage{hyperref}
\graphicspath{{figures/}}   
\usepackage{placeins}
\usepackage{epstopdf}

\usepackage{braket}
\usepackage{siunitx}
\usepackage{changes}
\usepackage{textcomp}

\binoppenalty=\maxdimen
\relpenalty=\maxdimen

\usepackage[plain]{fancyref}

\begin{document}

\title{Introducing coherent time control to cavity-magnon-polariton modes}

\author{Tim Wolz}	
\email{tim.wolz@kit.edu}
\author{Alexander Stehli}		
\author{Andre Schneider}
	\affiliation{Institute of Physics, Karlsruhe Institute of Technology, 76131 Karlsruhe, Germany}
\author{Isabella Boventer}
	\affiliation{Institute of Physics, Karlsruhe Institute of Technology, 76131 Karlsruhe, Germany}
	\affiliation{Institute of Physics, Johannes Gutenberg University Mainz, 55099 Mainz, Germany}
\author{Rair Mac\^edo}		
	\affiliation{James Watt School of Engineering, Electronics \& Nanoscale Engineering Division, University of Glasgow, Glasgow G12 8QQ, United Kingdom}
\author{Alexey V. Ustinov}	
	\affiliation{Institute of Physics, Karlsruhe Institute of Technology, 76131 Karlsruhe, Germany}
	\affiliation{Russian Quantum Center, National University of Science and Technology MISIS, 119049 Moscow, Russia}
\author{Mathias Kl\"aui}
	\affiliation{Institute of Physics, Johannes Gutenberg University Mainz, 55099 Mainz, Germany}
\author{Martin Weides}		
	\email{martin.weides@glasgow.ac.uk}
	\affiliation{Institute of Physics, Karlsruhe Institute of Technology, 76131 Karlsruhe, Germany}
	\affiliation{James Watt School of Engineering, Electronics \& Nanoscale Engineering Division, University of Glasgow, Glasgow G12 8QQ, United Kingdom}

\date{\today}


\begin{abstract}
By connecting light to magnetism, cavity-magnon-polaritons (CMPs) can build links from quantum computation to spintronics. As a consequence, CMP-based information processing devices have thrived over the last five years, but almost exclusively been investigated with single-tone spectroscopy. However, universal computing applications will require a dynamic control of the CMP on demand and within nanoseconds. In this work, we perform fast manipulations of the different CMP modes with independent but coherent pulses to the cavity and magnon system. We change the state of the CMP from the energy exchanging beat mode to its normal modes and further demonstrate two fundamental examples of coherent manipulation: First, a dynamic control over the appearance of magnon-Rabi oscillations, i.e., energy exchange, and second, a complete energy extraction by applying an anti-phase drive to the magnon. Our results show a promising approach to control different building blocks for a quantum internet and pave the way for further magnon-based quantum computing research.    
\end{abstract}


\maketitle

\section*{\label{sec:introduction}{Introduction}}
The cavity-magnon-polariton (CMP)~\cite{huebl_high_2013, zhang_strongly_2014, tabuchi_hybridizing_2014, goryachev_high-cooperativity_2014} is a hybrid particle arising from strong coupling between photon and magnon excitations. It interconnects light with magnetism being an excellent candidate to combine quantum information with spintronics~\cite{lachance-quirion_hybrid_2019, karenowska_magnon_2016}. The first CMP-based devices, such as a gradient memory~\cite{zhang_magnon_2015} and radio-frequency-to-optical transducers~\cite{hisatomi_bidirectional_2016} have already been developed. Especially the latter ones are crucial devices for a quantum internet, for instance, because they bridge microwave-frequency based quantum processors to long range optical quantum networks. Since the recent emergence of this hybrid particle, three different models, in particular, have helped to unravel the physics of CMPs over the last years: first, the picture of two coupled oscillators, which is the most intuitive one; the underlying physics, however, is only revealed from an electromagnetic viewpoint, which is the second model and shows a phase correlation between cavity and magnon excitation~\cite{bai_spin_2015}; and finally, the quantum description of the CMP, which has, for instance, given the theoretical framework for a coupling of magnons to a superconducting qubit~\cite{tabuchi_coherent_2015, lachance-quirion_resolving_2017}. Many spectroscopic experiments have led to new insights about loss channels~\cite{tabuchi_hybridizing_2014, kosen_microwave_2019, pfirrmann_magnons_2019}, their temperature dependence~\cite{boventer_complex_2018, zhang_cavity_2015, golovchanskiy_interplay_2019}, and to the observation of level attraction~\cite{grigoryan_synchronized_2018, harder_level_2018, boventer_control_2019}. These spectroscopic measurements, however, are performed under continuous driving, and while they have yielded great physical insight into these hybrid systems, flexible and universal information processing requires the manipulation of such physical states on demand and on nanosecond timescales. Despite this necessity for fast manipulation, the literature about time resolved experiments with either an yttrium iron garnet (YIG) waveguide~\cite{van_loo_time-resolved_2018} or CMPs~\cite{zhang_strongly_2014, zhang_magnon_2015, morris_strong_2017, match_transient_2019-1} is scarce and confined to cavity-pulsing. A simultaneous and coherent control over both subsystems has yet to be demonstrated, which is the subject of this work. We establish the control over the cavity \textit{and} magnon system by using coherent manipulation pulses on the timescale of nanoseconds. We observe the transition from maximum energy exchange to no energy exchange between the two quasi-particles depending on the applied pulses. Furthermore, we employ these results for a dynamic control of the different modes and for the extraction of the total energy from the system by destructive interference within the sample.

In our experiments the electromagnetic resonance of a copper cavity interacts with the Kittel mode - the uniform ferromagnetic resonance (FMR)~\cite{kittel_theory_1948} - of a YIG-sphere mounted inside the cavity. The Landau-Lifshitz-Gilbert (LLG) equation~\cite{landau_theory_1935}  describes the Kittel-mode as a macrospin with dynamic magnetization $m(t) = m \rm{e}^{-\rm{i} \mathit{\omega t}}$ in an external magnetic field $\bm{H}$. The cavity resonance can be modeled as an RLC circuit. Following Ref.~\cite{bai_spin_2015}, a linear coupling between both systems arises from their mutual back actions, leading to a phase correlation. The changing magnetization of the FMR induces an electric field in the cavity according to Faraday's law. Following Amp\`{e}re's law, the cavity field gives rise to a cavity current, which produces a magnetic AC-field $h(t) = h \rm{e}^{-\rm{i} \mathit{\omega t}}$ driving the FMR. Combining the LLG, the RLC equation and Maxwell's laws yields a system of coupled equations for $h$ and $m$ (Supplementary). If both subsystems are close to resonance, these equations can be simplified to the eigenvalue equations of two coupled harmonic oscillators with constant coupling strength~$g$:
\begin{equation}
\left(\begin{array}{cc}
\omega-\tilde{\omega}_{{\rm c}} & g\\
g & \omega-\tilde{\omega}_{\rm r}
\end{array}\right)\left(\begin{array}{c}
h\\
m
\end{array}\right)=0.
\label{eq:coupled_pendulums}
\end{equation}
Here, $\tilde{\omega}_{{\rm c}}$ and $\tilde{\omega}_{\rm r}$ denote the complex eigenfrequencies of the cavity and magnon system, respectively. They are defined as $\tilde{\omega}_{{\rm c}} = \omega_{\rm c} - \rm{i}\beta \omega_{\rm c}$ and $\tilde{\omega}_{\rm r} = \omega_{\rm r} - \rm{i}\alpha \omega_{\rm c}$ with bare cavity frequency $\omega_{{\rm c}}=1/\sqrt{LC}$, bare magnon frequency $\omega_{{\rm r}}=\gamma\sqrt{|H|(|H|+M_0)}$, where $\alpha$, $\beta$ are damping factors, $\gamma$ the gyromagnetic ratio of the Kittel-mode and $M_0$ its saturation magnetization. If both subsystems are exactly on resonance, i.e, at their crossing point, $\omega_{{\rm c}}=\omega_{\rm r}=\omega_0$, the eigenfrequencies of Eq.~(\ref{eq:coupled_pendulums}) are given by $\omega_{\pm}=\omega_0\pm g$ with the eigenvectors $\xi_{\pm}=\left(\begin{array}{cc}
1, & \pm1\end{array}\right)$, the so-called normal modes.
A single, short pulse to the cavity prepares the system in the non-eigenstate $\xi_{0, \rm{c}} = \xi_+ + \xi_- = \left(\begin{array}{cc}
1, & 0 \end{array}\right)$, known as beat mode. The excitation, and therefore the energy, periodically oscillates between cavity and magnon. Hence, the system displays classical magnon-Rabi oscillations~\cite{zhang_strongly_2014, match_transient_2019-1}. Figures.~\ref{fig:setup}\mbox{c-e} illustrate these different modes in the intuitive picture of two coupled pendula. To observe the normal modes $\xi_{\pm}$, where no energy is exchanged, one has to coherently and simultaneously excite the cavity and magnon system while recording the cavity response (Fig.~\ref{fig:setup}a).

\section*{Results}
\label{sec:sample}
\begin{figure}[tb]
	\includegraphics[width=\columnwidth]{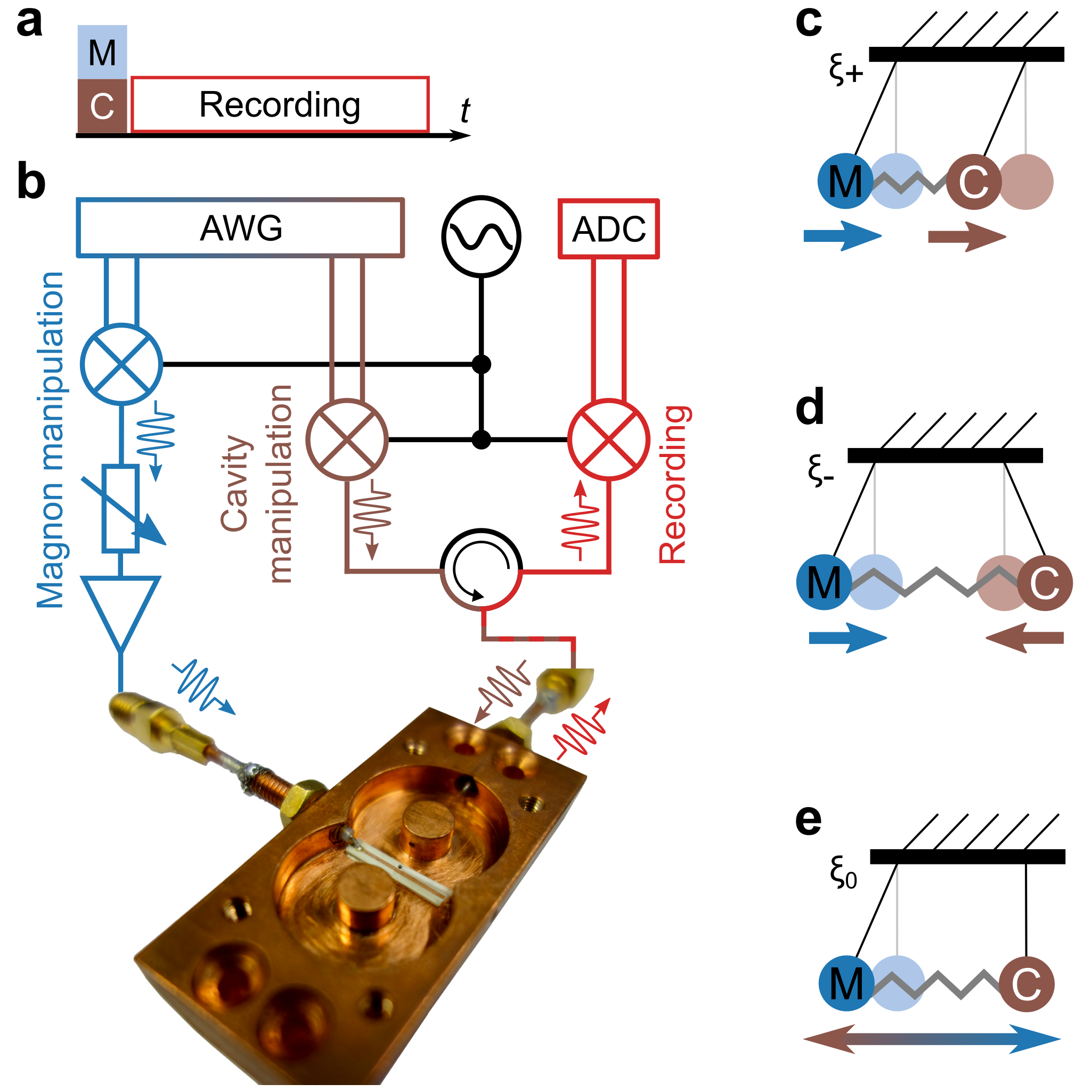}
	\caption{\textbf{Experimental setup and mode visualization.} \textbf{a} Typical pulse sequence used to prepare the system in its normal mode. \textbf{b} The time domain setup comprises the following three parts: the magnon manipulation line, the cavity manipulation line, and the recording line for the cavity response. A continuous signal of the microwave source is up-converted with pulses from the AWG, which then excites cavity and magnon system. The reflected and down-converted signal from the cavity is recorded by an ADC-card.
	\textbf{c-e} Pendula representation of the different CMP modes: in-phase mode $\xi_{+}$, anti-phase mode $\xi_{-}$, and beat-mode $\xi_{0}$.
	}
	\label{fig:setup}
\end{figure}

\label{sec:res}
Our sample is a copper reentrant cavity~\cite{goryachev_high-cooperativity_2014} resonating at $\omega_{r}/2\pi=\SI{6.58}{\giga \hertz}$. An additional stripline with a second microwave port is fixed to the bottom of the cavity. This port allows for the direct manipulation of the magnon mode in a YIG sphere with a diameter of \SI{0.5}{\milli \meter}. The sphere is placed close to the magnetic antinode of the cavity. The cavity's magnetic AC-field, the stripline AC-field and the external bias field stand all perpendicular to each other (Supplementary), which minimizes unwanted crosstalk between the two AC-fields. Measurements are performed with a time-domain setup (Fig.~\ref{fig:setup}b) comprising three parts: magnon manipulation, cavity manipulation and recording. It enables us to independently but coherently pulse the two subsystems and record the reflected signal, i.e., the outgoing photons, of the cavity. An arbitrary phase offset between all pulses can be chosen. Additionally, the applied power to the magnon can be adjusted allowing for an amplitude matching and thus equal excitation of cavity and magnon system.

\begin{figure}[tb]
	\includegraphics[width=\columnwidth]{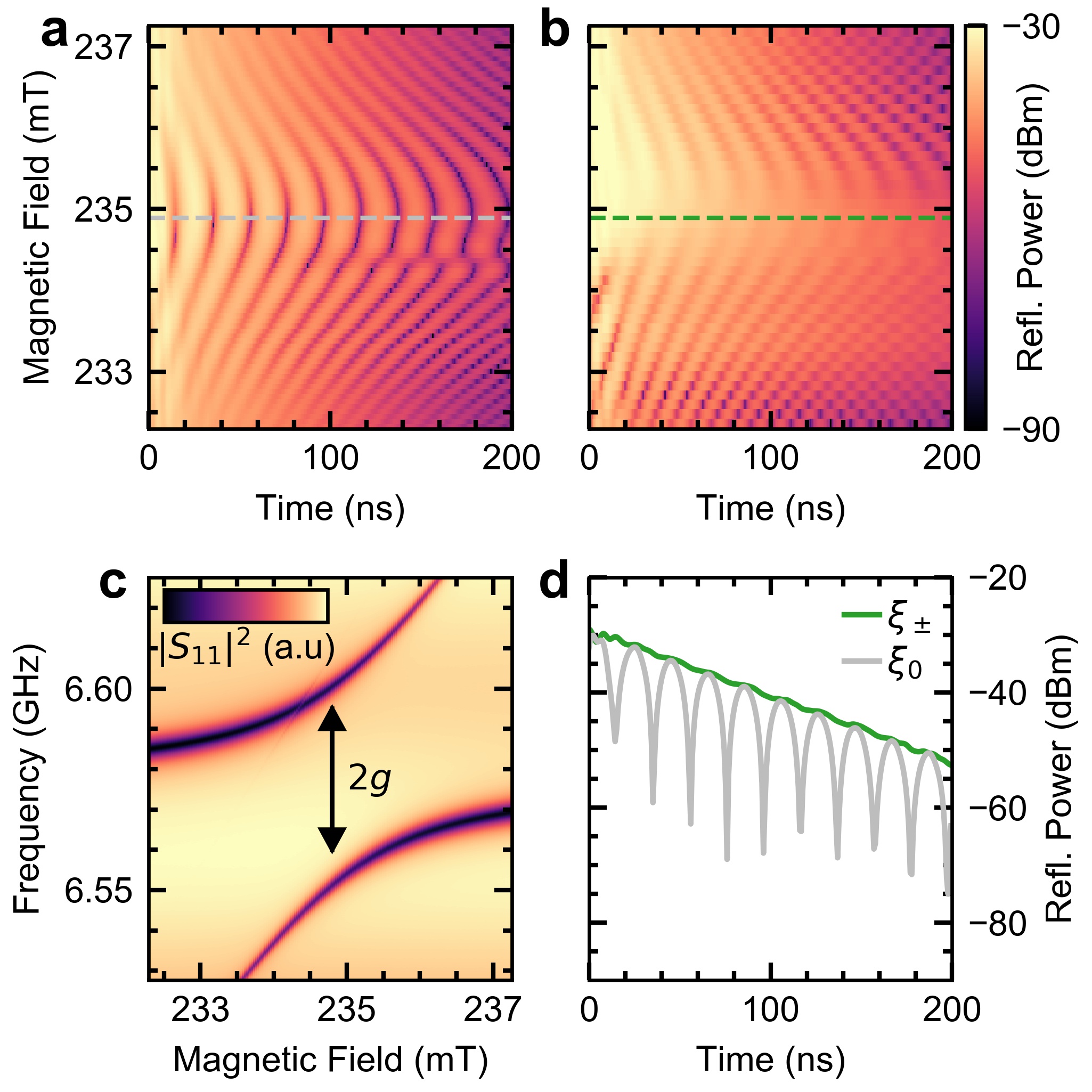}
	\caption{\textbf{Time-resolved and spectroscopic cavity response for the different CMP modes.} \textbf{a} Time evolution of the reflected cavity signal revealing magnon-Rabi oscillations after a single pulse to the cavity. Between each recorded time trace the external field is swept. Close to \SI{234}{\milli \tesla} another spurious mode is visible, particularly between \SI{100}{\nano \second} and \SI{200}{\nano \second}. \textbf{b} Cavity time evolution after phase and amplitude matched pulses to both cavity and magnon. Rabi oscillations on resonance (green dashed line) are suppressed since the system is prepared in one normal mode. \textbf{c} Avoided level crossing of the CMP, probed spectroscopically. \textbf{d} Comparison of the cavity's time evolution in one normal mode $\xi_{\pm}$ (green line) and beat mode $\xi_{0}$ (gray line) at the crossing point. Time traces are line cuts along the dashed lines in \textbf{a} and \textbf{b}.
	}
	\label{fig:chevrons}
\end{figure}

The avoided level crossing data (Fig.~\ref{fig:chevrons}c), measured spectroscopically, shows a coupling strength of $g/2\pi=\SI{24.6}{\mega \hertz}$, as theoretically expected for this cavity-magnon-system~\cite{boventer_complex_2018}, and identical decay rates (Supplementary) at the crossing point of $\kappa_{\rm{crp}}/2\pi = \SI{2.1}{\mega \hertz}$ due to equal hybridization. We hence conclude that our system is strongly coupled. This result is also validated in the time domain (Fig.~\ref{fig:chevrons}a). The external field, and therefore $\omega_{{\rm r}}$, is swept and the cavity is excited with a single short pulse in between the sweep steps. The reflected signal shows clear Rabi-oscillations confirming the coupling strength of $g/2\pi=\SI{24.6}{\mega \hertz}$ and thus exhibiting an oscillation period of $t_{\rm R}=2\pi/g=\SI{40.6}{\nano \second}$ for $\omega_{{\rm c}}=\omega_{{\rm r}}$. The measured decay time of $\tau = \SI{77.6}{\nano \second}$ is also in good agreement with $1/\kappa_{\rm{crp}} = \SI{75.8}{\nano \second}$. Both, time resolved and spectroscopic data exhibit another weakly coupled magnon-mode at around \SI{234}{\milli \tesla}, which slightly distorts the signal of the pure Kittel mode but is not of interest for our experiments.

\begin{figure}[tb]
	\includegraphics[width=\columnwidth]{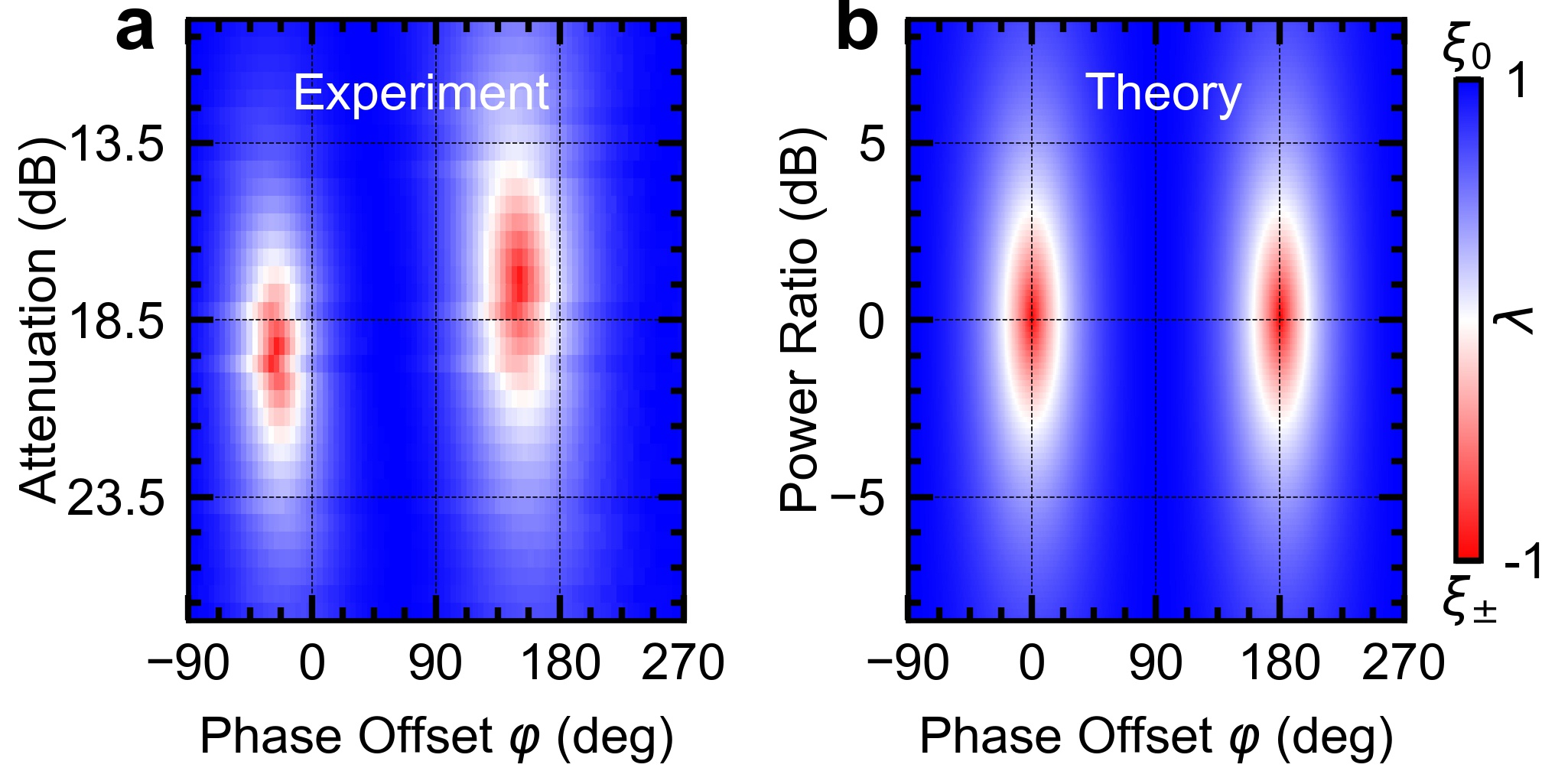}
	\caption{\textbf{Mode composition of the CMP} depending on applied power ratio and phase offset $\varphi$ between the two pulses. Experimental data of the cavity response (\textbf{a}) are fitted to Eq.~(\ref{eq:timetrace}) and can then be compared to analytic data (\textbf{b}) . The chosen attenuation in the magnon line (y-axis of \textbf{a}) corresponds to the power ratio of the drive pulses used in \textbf{b} with an experimentally inaccessible offset.  The parameter $\lambda$ translates to the mode composition of the CMP. Within the red ellipses, the CMP is predominantly excited in its normal modes. Experimentally found normal mode ellipses are slightly shifted to lower phase values and differ in power ratio compared to the simulated data due to a minimal timing mismatch of the applied pulses, direct crosstalk or small frequency drifts in the system (Supplementary).
	}
	\label{fig:trans_modes}
\end{figure}

After the characterization of our system, we apply an additional pulse directly to the magnon system. The cavity response of such a two-pulse experiment is shown in Fig.~\ref{fig:chevrons}b. The two pulses are phase and amplitude-matched for the on-resonance-case in order to prepare the system in its normal mode. Since no energy is exchanged in the normal modes, a pure exponential decay is expected and observed at the crossing point. However, if $\omega_{{\rm c}} \neq \omega_{{\rm r}}$, the amplitude and phase-matching does not hold and the Rabi oscillations are visible. But when the system approaches its crossing point (green dashed line in Fig.~\ref{fig:chevrons}b), the dips of the oscillations become more shallow than in Fig.~\ref{fig:chevrons}a, until they are almost completely suppressed. The slight remaining oscillations left are due to experimental imperfections. Figure~\ref{fig:chevrons}d shows a line cut at the crossing point of the single and two-pulse experiment emphasizing the different responses of the normal mode and beat mode. The two-pulse response reveals the expected exponential decay of either one of the normal modes $\xi_{\pm}$. Following the external drive pulses, the cavity and magnon field have the same amplitude and oscillate in-phase (anti-phase) for $\xi_{+}$ ($\xi_{-}$), which is characteristic for the normal modes~\cite{bai_spin_2015}.

We also monitor the transition from $\xi_0$ to $\xi_{\pm}$ by sweeping the phase offset and amplitude ratio between the two pulses at the crossing point. The cavity response during free evolution is recorded for every set of phase-offset $\varphi$ and applied power ratio, and fitted to the following formula (See supplementary for details on the derivation):
\begin{equation}
P_{{\rm c}}(t)=p_0\left[(1-\lambda)+(1+\lambda)\cos^{2}(g\,t+\phi_0)\right]{\rm e}^{-t/\tau},
\label{eq:timetrace}
\end{equation}
which describes the cavity response during free evolution. The parameter $\lambda$ has inherent bounds of $-1$ to $1$ and determines the behavior of the system. $\lambda=1$ gives a damped sine function and thus represents the beat mode $\xi_{0}$, whereas $\lambda=-1$ yields the pure exponential decay of the normal modes $\xi_{\pm}$. A proportionality constant $p_0$ normalizes the different input powers, $\phi_0$ describes the initial phase of the beating and $\tau$ the decay time.  Figure~\ref{fig:trans_modes} displays the extracted values for $\lambda$ corresponding to the different modes for the measured data and can be compared to the analytic solution. As expected from the oscillator and the electromagnetic model~\cite{bai_spin_2015}, where the phase between magnon field and cavity field at the crossing point is locked to either in phase or complete anti-phase for the two eigenmodes, the system is prepared in the normal modes (red regions) for matching powers and phase offsets of 180\textdegree \@ and \@ 360\textdegree. In the experiment, the red regions are shifted to lower phase offset values by roughly 30\textdegree. This phase shift translates to a timing mismatch between the two applied pulses below \SI{0.1}{\nano \second}, which is beyond the precision of our setup. The slope between the normal mode regions is either due to crosstalk or little drifts of the external magnetic field. We have verified the reasons for these deviations from the exact analytic solution by numerical simulations of two coupled oscillators with short drive pulses (Supplementary). Apart from these little discrepancies, which are purely limitations of the experimental setup, our collected data agrees well with theory. We can change the parameter $\lambda$ and therefore the CMP mode composition continuously. With a second pulse, the phase relation between cavity current and magnon magnetization can be set to an arbitrary value. Thus, these results extend the work of Bai \textit{et.\@ al}, where the phase relation between the two systems is fixed by the external field~\cite{bai_spin_2015}. This pulsed mode control of the CMP may hence benefit future spin rectification~\cite{harder_electrical_2016} experiments and applications.

\begin{figure}[tb]
	\centering
	\includegraphics[width=\columnwidth]{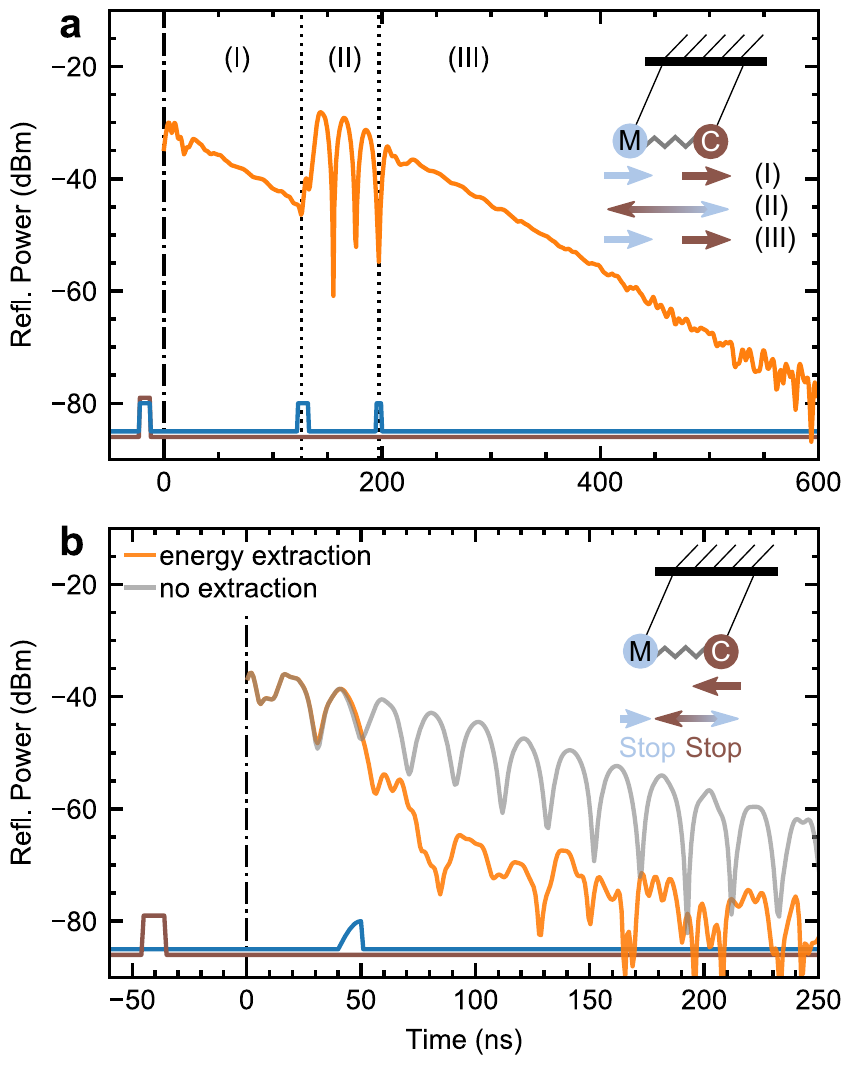}
	\caption{
		\textbf{Coherent and dynamic control over the CMP.} \textbf{a} Time trace of the cavity response showing dynamic control over the time span of energy exchange.  The time trace is divided into three parts: (I) The system is in its normal mode with almost no energy exchange; (II) An additional pulse to the magnon system introduces an energy difference leading to Rabi oscillations; (III) A third pulse to the magnon extracts energy out of the system by destructive interference, bringing it back to the normal mode. \textbf{b} Cavity time trace showing energy extraction by an anti-phase drive of the magnon, which counters the energy coming from the cavity. Photons interfere destructively and the CMP is completely deexcited by destructive interference after the magnon pulse. Solid brown and blue lines represent the applied pulse sequences for the cavity and magnon system, respectively (pulse-height not scaled). The ring up of cavity and magnon system, $t<\SI{0}{\nano \second}$, is omitted for clarity. Insets show the corresponding coupled pendula visualizations.  
	}
	\label{fig:coherence_control}
\end{figure}

The external control of the CMP mode composition, which we showed and described, is directly linked to the amplitude control of Rabi oscillation and thus to the amount of energy transferred between the two subsystems. Seizing this opportunity, we now demonstrate a coherent and also dynamic control over the CMP during one single decay by choosing an arbitrary period in which the magnon-Rabi oscillations are allowed to occur (Fig.~\ref{fig:coherence_control}a). Having prepared the system on resonance, we excite both magnon and cavity with phase and amplitude-matched pulses, to bring the CMP into its normal mode $\xi_{\pm}$ and observe a pure exponential decay. We then increase the energy of the magnon subsystem by pulsing it again with a short pulse. The whole system is now in a superposition of $\xi_{+}$ and $\xi_{-}$, i.e., in its beat mode. Rabi oscillations are visible and energy is exchanged. After a few oscillations, a third pulse in anti-phase to the incoming photons from the cavity and with lower amplitude, due to energy loss in the system, extracts the additional energy, previously introduced to the magnon subsystem, by destructive interference and brings the whole system back to its normal mode. The Rabi oscillations stop and a simple exponential decay is visible, again. In the picture of two coupled pendula the three different segments of the decay corresponds to (I) both pendula oscillating in phase, (II) a strong drive of one pendulum introduces energy leading to the beat mode, and (III) and a careful short deacceleration brings the system back to the normal mode. 

In a second experiment (Fig.~\ref{fig:coherence_control}b), we apply the technique used for active noise control~\cite{elliott_active_1990, kuo_active_1999} in acoustics to the CMP: The cavity is excited by a short pulse and the energy is transferred to the magnon system and back to the cavity. During the second energy transfer to the magnon, we drive the magnon in an anti-phase manner to the oscillation of the incoming photons. A destructive interference extracts all the stored energy from the system and thus the reflected power of the cavity drops within a few nanoseconds by roughly \SI{20}{\decibel}, before the signal reaches the baseline of the measurement setup. This behavior can also be understood intuitively in the picture of the coupled pendula: A first pulse tilts only one pendulum, the energy is transferred with time and the second pendulum starts oscillating. Exactly in the moment when all energy is transferred, i.e., the first pendulum is at rest, the other pendulum is stopped abruptly by the external second pulse and the whole system is deexcited. These two experiments presented here demonstrate the fundamentals of dynamic and coherent control over the CMP.

\section*{Discussion}
We presented coherent time-domain control of both cavity and magnon while recording the cavity response, as well as real-time manipulation of the CMP. We also showed the transition from the beat mode to the normal modes of the CMP and explained it with the theory models provided in Ref.~\cite{bai_spin_2015}. The CMP can be set in an arbitrary superposition of its eigenmodes depending on the phase offset and amplitude of the applied pulses. This pulse influence agrees well with theory considering the finite time resolution of our setup. Furthermore, we demonstrated a coherent control over the CMP, with which the amount of transferred energy as well as the total amount of energy in the system can be manipulated at any given time. 

Spectroscopic two-tone experiments predicted and observed the regime of level attraction~\cite{zhang_observation_2017, grigoryan_synchronized_2018, boventer_complex_2018}, which has also been linked to an entanglement of photon and magnon~\cite{yuan_steady_2019}. Our technique would allow to prepare the CMP in this regime and then observe its time-evolution. Although our demonstration was purely classical, the presented control can readily be applied at cold temperatures, i.e., in the single magnon regime and the predicted entanglement in the level attraction regime may be verified. Moreover, with the demonstration of dynamic and coherent control, we have added another instrument to the toolbox for the construction of a quantum internet~\cite{kimble_quantum_2008}. Together with magnon based storage~\cite{zhang_magnon_2015} and qubit magnon coupling~\cite{tabuchi_coherent_2015, lachance-quirion_hybrid_2019}, we believe that our work will advance the encoding of qubit / Fock states in magnons~\cite{rezende_coherent_1969} similar to superconducting resonators~\cite{hofheinz_generation_2008, leghtas_deterministic_2013} and subsequently the implementation of bosonic gates~\cite{vlastakis_deterministically_2013, heeres_implementing_2017}. The CMP's significant potential as an interface from radio frequency to optics~\cite{hisatomi_bidirectional_2016} balances the short lifetime of magnons compared to superconducting resonators. Thus, our results promise a link between the different building blocks for a quantum network and open new ways for magnon based quantum computation research.

Finally, our work demonstrates the fundamental principle of time-control of the individual components in hybrid systems. Applied to other compound devices featuring polaritons from the strong coupling of electromagnetic waves with electric or magnet excitations, such as optomechanics~\cite{aspelmeyer_cavity_2014} or electromechanics~\cite{regal_cavity_2011}, it provides a flexible platform that intrigues fundamental coherent control of the strong light-matter interaction dynamics. 

\section*{Methods}
\subsection*{Experimental setup}
The experimental setup is adapted from  quantum simulation experiments with superconducting qubits~\cite{braumuller_analog_2017}. Its core components are a microwave source, an arbitrary waveform generator (AWG) with two sets of DACs, and a two-channel ADC-card. For our experiments it is vital that the phase between magnon and cavity control pulses is independently controllable but also stable over the entire experiment. We ensure this by using a single microwave source and two DAC sets in combination with the internal clock of the AWG for both DAC sets. The continuous signal generated by the microwave source is up-converted to $\omega_0$ via separate but identical IQ mixers and \SI{10}{\nano \second} short IQ-pulses with a carrier frequency of \SI{250}{\mega \hertz} from the AWG. The up-conversion preserves the phase offset and the envelope of the IQ-pulses emitted by the AWG. A voltage controllable attenuator in combination with a \SI{26}{\decibel} amplifier inserted in the magnon line enables us to adjust the excitation amplitude and hence vary the power ratio between magnon and cavity excitation pulses.

\subsection*{Data acquisition}
The cavity response is recorded by measuring the IQ components of the down converted, filtered and amplified signal with the ADC-card. A subsequent digital down conversion removes the \SI{250}{\mega \hertz} carrier frequency and yields amplitude and phase data. All data acquisition and analysis are performed with the open source measurement suite qkit \url{https://github.com/qkitgroup/qkit}.

\subsection*{Experimental technique}
The initial two pulses for the cavity and magnon system have to reach the sample simultaneously in order to ensure a good phase and amplitude matching (Supplementary). We therefore calibrate the cable delay between the two input lines by emitting two Gaussian shaped pulses simultaneously at the AWG, which are sent to both subsystems. Due to undeterrable crosstalk, a part of the pulse applied to magnon system is transferred to the recording line of the cavity. Recording the reflected cavity pulse and the transmitted magnon pulse, we find a cable delay of $\SI{7}{ns}$ by fitting the pulses and extracting their mean values. However due to simplicity, square pulses are used for most experiments.
Because of inaccessible and fluctuating parameters, such as uncorrectable cable delays below \SI{1}{\nano \second}, drifts in the external fields and unknown reflected parts of the emitted pulses, the correct phase offsets and power ratio for the specific experiments are found experimentally by a sweep of these two parameters. Although the eigenfrequencies are shifted by $g$ compared to the bare resonator frequency, the system is always pulsed at $\omega_0$. This gives the best experimental compromise for equally exciting the different modes of the system. All experiments are performed in the linear regime (Supplementary).

\subsection*{Sample details}
The employed YIG-sphere is commercially available from Ferrisphere Inc. The stripline is \SI{50}{\Omega} matched and open-ended. It is made from a Rogers TMM10i copper cladded (\SI{35}{\micro \meter}) substrate with a thickness of \SI{0.64}{\milli \meter}.

\begin{acknowledgments}
We acknowledge valuable discussions with Konrad Dapper, Bimu Yao and Can-Ming Hu.
This work was supported by the European Research Council (ERC) under the Grant Agreement 648011, Deutsche Forschungsgemeinschaft (DFG) within Project No.\@ WE4359/7-1 and INST 121384/138-1, through  SFB  TRR  173/Spin+X, and the Initiative and Networking Fund of the Helmholtz Association.
T.W.\@ acknowledges financial support by Helmholtz  International  Research  School  for  Teratronics (HIRST), A.St.\@ by the Landesgraduiertenf\"orderung (LGF) of the federal state Baden-W\"urttemberg, A.Sch.\@ by the Carl-Zeiss-Foundation and R.M.\@ by the Leverhulme Trust. A.V.U.\@ acknowledges partial support from the Ministry of Education and Science of the Russian Federation in the framework of the contract No.\@ K2-2017-081.
\FloatBarrier
\end{acknowledgments}

\section*{Author contributions}
T.W. and M.W. conceived the experiment. T.W. performed the measurements with support by A.St., A.Sch., and I.B. T.W. carried out data analysis with contributions from A.St. and R.M.  T.W. wrote the manuscript with input from and discussions with all co-authors. A.V.U., M.K, and M.W. supervised the project.

\bibliography{FMR_TD2}

\clearpage
\onecolumngrid
\appendix

\renewcommand{\thefigure}{S\arabic{figure}}
\renewcommand{\theequation}{S\arabic{equation}}
\setcounter{figure}{0}
\setcounter{equation}{0}
\section*{Supplementary Material}
\FloatBarrier
\subsection{Visualization of the CMP coupling mechanism}
\begin{figure}[tbp]
	\includegraphics{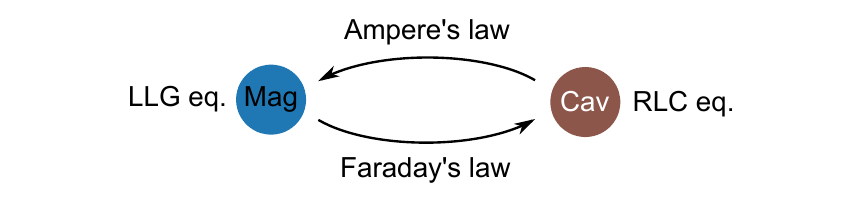}
	\caption{\textbf{Illustration of the electrodynamic CMP  model} and its coupling mechanisms, according to Bai \textit{et~al.}\cite{bai_spin_2015}.}
	\label{fig:pendulums}
	\centering
\end{figure}

According to Ref.~\cite{bai_spin_2015} four equations govern the behavior of the CMP: (i) the LLG equation describes the dynamics of the macrospin representing the FMR; (ii) the RLC equation gives the current dynamics of the cavity; (iii) Amp\`{e}re's law shows the coupling from cavity to magnon because the cavity current drives the magnetization of the FMR; (iv) a time dependent change in the magnetization produces a voltage acting on the cavity, according to Faraday's law. Fig.~\ref{fig:pendulums} illustrates this model. Combining all these equations, one obtains the following coupled system of equations:
\begin{equation}
\left(\begin{array}{cc}
\omega^{2}-\omega_{\rm c}^{2}+2{\rm i}\beta\omega_{\rm c}\omega & {\rm i}\omega^{2}K_{\rm c}\\
-{\rm i}\omega_{\rm m}K_{\rm m} & \omega-\omega_{\rm{r}}+{\rm i}\alpha\omega
\end{array}\right)\left(\begin{array}{c}
j\\
m
\end{array}\right)=0.
\label{eq:coupled_cmp_electro}
\end{equation}
Here, $K_{\rm m}$ and $K_{\rm c}$ are coupling constants, $\omega_m = \gamma M_0$ and all other variables as defined in the main text. This model describes the discussed phase correlation. On resonance, the equations can be simplified to the coupled oscillator model, given by Eq.~(\ref{eq:coupled_pendulums}) in the main text.

\subsection{Field arrangement inside the cavity}
\begin{figure}[tbp]
	\includegraphics{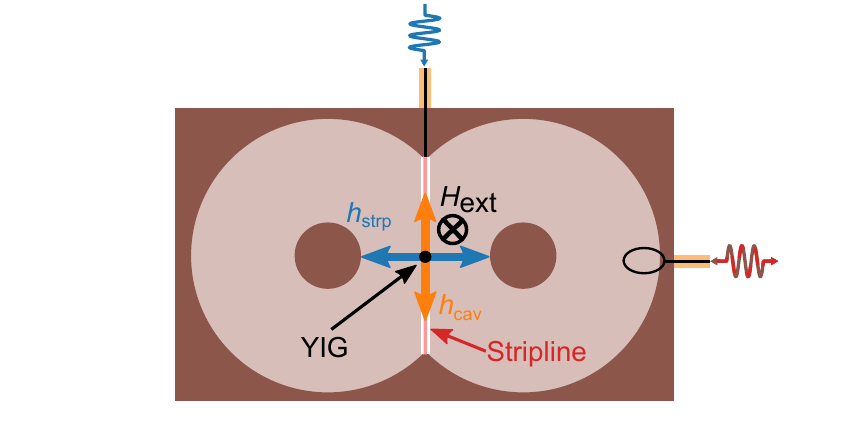}
	\caption{\textbf{Field arrangement inside the reentrant cavity.} The external magnetic field $H_{\rm{ext}}$ and the two AC-fields, $h_{\rm{strp}}$ and $h_{\rm{cav}}$, are aligned perpendicular towards each other. Signals are coupled into the cavity via an inductive loop and into the magnon system via a stripline.}
	\label{fig:fields}
	\centering
\end{figure}

The magnetic field arrangement is depicted in Fig.~\ref{fig:fields}. The external field penetrating the cavity is aligned in parallel to the two posts inside the cavity. The magnetic field of the cavity mode circulates around the post and interferes constructively in the middle between the posts. Here, the magnetic AC-field of the cavity is aligned parallel to the stripline. According to Amp\`{e}re's law, the stripline's AC-field circulates around the stripline, giving a perpendicular orientation towards the other fields.

\subsection{Line widths and decay times}
\begin{figure}[tbp]
	\includegraphics{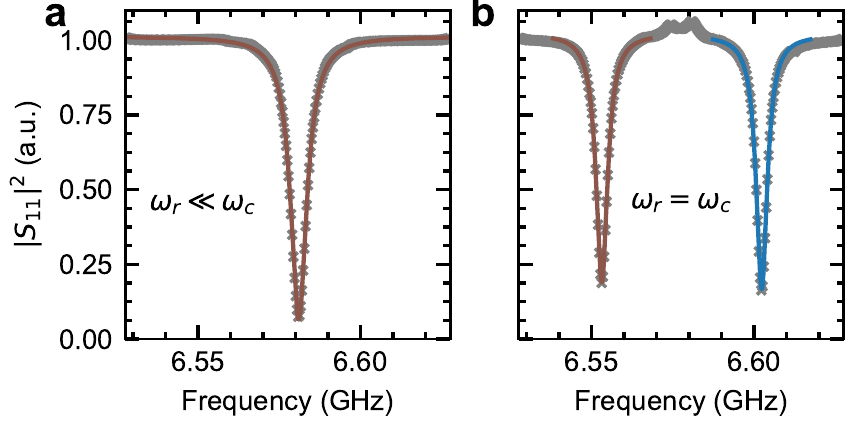}
	\caption{\textbf{Spectroscopic linewidth measurements} of \textbf{a} cavity detuned from the crossing point and \textbf{b} cavity and magnon system at the crossing points. The solid lines represent Lorentzian fits to the background corrected measurement data. The small peak in the baseline in b) is due to the background correction and occurrence of another mode, as described in the main text.}
	\centering
	\label{fig:dec_times}
\end{figure}
We use a Lorentzian fit to extract the off-resonant linewidth (HWHM), i.e., decay rate of the cavity from the avoided level crossing data with the magnon detuned (Fig.~\ref{fig:dec_times}a) and find $\kappa_{\rm{c}}/2\pi = \SI{3.0}{\mega \hertz}$ corresponding to $\beta=\kappa_{\rm{c}}/\omega_{\rm{c}}=\SI{0.046}{\percent}$ ($Q_L=1090$). Tuning in the magnon, the linewidth of the cavity decreases as expected~\cite{bai_spin_2015} until, at the crossing point, both dips hybridize with equal amounts (Fig.~\ref{fig:dec_times}b) leading to decay rates of $\kappa_{\rm{crp}}/2\pi = \SI{2.1}{\mega \hertz}$ corresponding to  $(\alpha + \beta)/2 = \SI{0.032}{\percent}$. The frequency independent Gilbert damping factor $\alpha$ of the magnon system can then be calculated as $\alpha = \SI{0.014}{\percent}$ ($Q=3570$).

\subsection{Estimation of the local magnetic field strength}
We explained our results from the viewpoint of two coupled linear oscillator. Here, we present a rough estimate verifying that the magnon system is indeed only driven linearly.  We use Ref.~\cite{gurevich_magnetization_1996} for the saturation field of a ferromagnetic resonance in a sphere, which is given by
\begin{equation}
\mu_0 H_{\rm{sat}} = 0.5\, \mu_0 \, \Delta H,
\end{equation}
where $\Delta H$ denotes the FWHM of the spin wave resonance. Detuned from the cavity resonance we find for our setup $\Delta \omega_r = \SI{1.5}{\mega \hertz}$. With the dispersion relation of the Kittel-mode, this value corresponds to $\mu_0 \Delta H \approx \SI{54}{\micro \tesla}$. Thus, the magnetic AC-field of our drive must not exceed half ot this value. The rf-power reaching the sample, i.e., the power emitted by the microwave source but then attenuated through cables and microwave components ranges from $\SI{-5}{\decibel m}$ to $\SI{+5}{\decibel m}$. The maximum power corresponds to a current of $I=\SI{8}{\milli \ampere}$. We now approximate the magnon transmission line as a cylindrical conductor and use Amp\`{e}re's law 
\begin{equation}
\mu_0 H=\frac{\mu_0 I}{2 \pi r},
\end{equation}
to calculate the AC-field strength with a distance $r=\SI{2}{\milli \meter}$. This gives a field strength of $\mu_0 H_{\rm{ac, max}}=\SI{5}{\micro \tesla}$ as an absolute maximum value and thus
\begin{equation}
\mu_0 H_{\rm{ac}} \leq \SI{5}{\micro \tesla} \ll \mu_0 H_{\rm{sat}} \approx \SI{27}{\micro \tesla}.
\end{equation}
Hence, we can conclude that all experiments were conducted in the linear regime.

\subsection{Time evolution of the cavity energy}
Eq.~(\ref{eq:timetrace}) in the main text describes the reflected power of the cavity, which is proportional to the stored energy inside the cavity. It is written in a form that the energy of the normal modes and beat mode occur in two separate terms with the parameter $\lambda$ giving the ratio of the two different modes. For our experiment, $\lambda$ should be a function of phase offset $\varphi$ and applied power ratio. At the crossing point, the two systems hybridize leading to same decay rates. Assuming similar coupling to the input ports, i.e., transmission lines, the applied power ratio equals the ratio of stored energy $\Delta^{2}=A_{\mathrm{m}}^{2}/A_{\mathrm{c}}^{2}$, where $A_{i}$ describes the maximum amplitude of the oscillation with only potential energy present. To derive Eq.~(\ref{eq:timetrace}) and find an analytic expression for $\lambda$, we start with the general solution for two coupled oscillators:
\begin{align}
	x_{{\rm c}}= & A\cos((\omega_{0}+g)\,t+\phi_0)+B\cos((\omega_{0}-g)\,t-\phi_0)\label{eq:cp_equ}\\
	x_{{\rm r}}= & A\cos((\omega_{0}+g)\,t+\phi_0)-B\cos((\omega_{0}-g)\,t-\phi_0).\label{eq:cp_equ2}
\end{align}
$A$, $B$ and $\phi_0$ are free parameters depending on the initial condition, i.e., the state of the oscillator when the pulses have stopped. Damping is neglected because the exponential decay can be factored out. The energy of the cavity oscillator is given by 
\begin{equation}
E_{{\rm c}}(t)=\frac{1}{2}\omega_{0}^{2}x_{{\rm c}}^{2}+\frac{1}{2}\dot{x}_{{\rm c}}^{2}. \label{eq:energy}
\end{equation}
Substituting Eq.~(\ref{eq:cp_equ}) and its derivative into Eq.~(\ref{eq:energy}) while neglecting terms proportional to $g$, since $g\ll\omega_{0}$, yields
\begin{equation}
E_{c}(t)=\frac{1}{2}\omega^{2}(\left|A\right|^{2}+\left|B\right|^{2})+\omega^{2}\left|AB\right|-2\omega^{2}\left|AB\right|\sin(g\,t+\phi_0)^{2}. \label{eq:energy_AB}
\end{equation}
Here, we have already denoted $A$ and $B$ as complex parameters with their absolute values. In a next step, we have to transform the initial conditions of the non-diagonal system, i.e., $\varphi$ and
$\Delta$ into $A$ and $B$:
\begin{equation}
\left(\begin{array}{c}
A\\
B
\end{array}\right)=\left(\begin{array}{cc}
1 & 1\\
1 & -1
\end{array}\right)\left(\begin{array}{c}
1\\
\Delta{\rm e^{{\rm i\varphi}}}
\end{array}\right)=\left(\begin{array}{c}
1+\Delta{\rm e^{{\rm i\varphi}}}\\
1-\Delta{\rm e^{{\rm i\varphi}}}
\end{array}\right).\label{eq:AB_vectors}
\end{equation}
Inserting these vectors into Eq.~(\ref{eq:energy_AB}) gives
\begin{align}
	E = & \frac{\omega_{0}^{2}}{2}\,\left(2\left(1+\Delta^{2}\right)+2\sqrt{\left(1-\Delta^{2}\right)^{2}+4\Delta\sin^2\varphi}-4\sqrt{\left(1-\Delta^{2}\right)^{2}+4\Delta\sin^2\varphi}\sin(g\,t+\phi_0)^{2}\right)\\
	E = & \frac{\omega_{0}^{2}}{2}\,\left(\underbrace{2\left(1+\Delta^{2}\right)-2\sqrt{\left(1-\Delta^{2}\right)^{2}+4\Delta\sin^2\varphi}}_{c(1-\lambda)}+\underbrace{4\sqrt{\left(1-\Delta^{2}\right)^{2}+4\Delta\sin^2\varphi}}_{c(1+\lambda)}\cos(g\,t+\phi_0)^{2}\right). \label{eq:E_complete}
\end{align}

Eq.~(\ref{eq:E_complete}) is now in the same form as Eq.~(\ref{eq:timetrace}). We can identify $(1-\lambda)$ and $(1+\lambda)$ and hence solve for $\lambda$ and $c$:
\begin{align}
	\label{eq:a_rabi}
	\lambda & = \frac{-\Delta^2+3\sqrt{\left(\Delta^2+1\right)^2 -  4\Delta^2\cos^2\varphi}-1}{\Delta^2+\sqrt{\left(\Delta^2+1\right)^2 - 4\Delta^2\cos^2\varphi}+1}\\
	c & = \Delta^2+\sqrt{\left(\Delta^2+1\right)^2 - 4\Delta^2\cos^2\varphi}+1.
\end{align}
The results of Eq.~(\ref{eq:a_rabi}) for different $\Delta^2$ and $\varphi$ are plotted in Fig.~\ref{fig:trans_modes}b in the main text.
\FloatBarrier
\subsection{Influence of experimental imperfections and numerical simulations}
\begin{figure}[tbp]
	\includegraphics{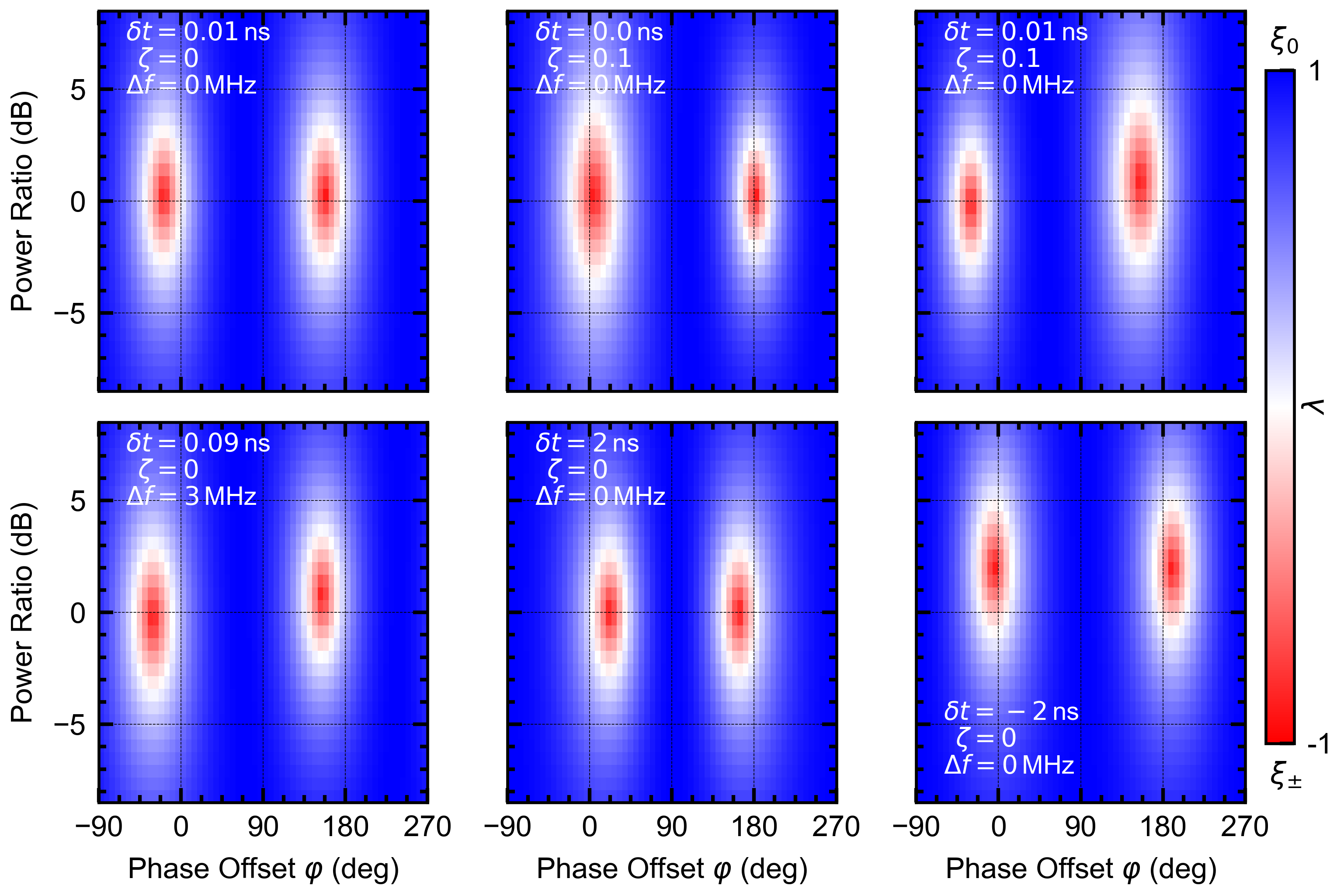}
	\caption{\textbf{Numerical simulated state of the CMP} and its influence of pulse delay $\delta t$, crosstalk $\zeta$, and detuning $\Delta f$. A slight timing mismatch of the two applied pulses leads to a phase shift of the normal modes. Detuning and/or crosstalk results in a slope between the normal mode regions. With increasing $\delta t$ the sinusoidal phase dependence becomes more asymmetric.}
	\centering
	\label{fig:delay_plots}
\end{figure}

Investigating the shifted ellipses found in the measurement values of Fig.~\ref{fig:trans_modes} in the main text, we
perform numerical simulations of two coupled oscillator with a short driving pulse. This gives us the possibility to test the influence of crosstalk, timing mismatches and a detuning between magnon and cavity. The equations of motions for two coupled oscillators, which lead to Eq.~(\ref{eq:coupled_pendulums}), and with drives included, are given by
\begin{align}
	\ddot{x}_{\rm c}+2\beta \omega_{\rm c}\dot{x}_{\rm c}-g^{2}\omega_{\rm c}x_{\rm r}& = (1-\zeta) F_{\rm c} \, + \, \zeta F_{\rm r} \label{eq:coupled_systems1} \\
	\ddot{x}_{\rm r}+2\alpha \omega_{\rm r}\dot{x}_{2}-g^{2}\omega_{\rm r}x_{\rm c}& = (1-\zeta)F_{\rm r} \, + \, \zeta F_{\rm c},
	\label{eq:coupled_systems2}
\end{align}
with
\begin{align}
	F_{\rm c}& = \cos(\omega_{0}t)\,\theta(t-t_{1,1})\,\theta(t_{1,2}-t), \\
	F_{\rm r}& = \Delta \cos(\omega_{0}(t-\delta t) - \varphi)\,\theta(t-t_{2,1})\,\theta(t_{2,2}-t).
\end{align}
$x_{\rm r}$ and $x_{\rm c}$ denote the amplitudes of the oscillators, $\alpha$ and $\beta$ the damping factors. The drives are represented by $F_1$ and $F_2$ with theta functions allowing for \SI{10}{\nano \second} short square pulses, as in our experiment.  A possible timing mismatch can be introduced by $\delta t = t_{2,1} - t_{1,1} = t_{2,2} - t_{1,2}$. The parameter $\zeta$ defines the amount of crosstalk. The theoretical case without any crosstalk in the system is thus given by $\zeta=0$. A possible detuning of $\Delta f = (\omega_{\rm r} - \omega_{\rm c})/2\pi$ can also be investigated. The cavity frequency and the drive frequency $\omega_0$ are kept constant. Eq.~(\ref{eq:coupled_systems1}) and Eq.~(\ref{eq:coupled_systems2}) are numerically solved with the scipy integrate package. We then use the resulting simulated time traces and their derivative to calculate $E(t)$ and fit it to Eq.~(\ref{eq:timetrace}) in the main text. The parameter $\lambda$ is plotted for different configurations of $\delta t$, $\Delta f$ and $\zeta$ in Fig.~\ref{fig:delay_plots}. It is evident that even a slight timing mismatch leads to a deviation from the ideal case, where the normal modes occur at exactly the same applied power to both systems and phase offset values of \SI{180}{\degree} and \SI{360}{\degree}. Instead, with increasing time delay $\delta t = t_{\rm{mag}} - t_{\rm{cav}}$, the phase interval between in-phase and anti-phase mode (starting with the in-phase mode) becomes greater than \SI{180}{\degree} and increases with  $\delta t$. For negative $\delta t$, the opposite behavior is the case.  One can observe a slope between the two elliptical normal mode regions if crosstalk is allowed. This is due to the phase depending constructive or destructive interference of the two pulses at the sample. A similar effect occurs if the magnetic field drifts and hence $\omega_{{\rm c}} \neq \omega_{{\rm r}}$. Then one system is closer to $\omega_{0}$ and receives therefore more energy than the other system. 

\end{document}